# Stochastic quantization weakening and quantum entanglement decoherence

Piero Chiarelli

*National Council of Research of Italy, Area of Pisa, 56124 Pisa, Moruzzi 1, Italy*

*Interdepartmental Center "E.Piaggio" University of Pisa*
Phone: +39-050-315-2359
Fax: +39-050-315-2166

Email: pchiare@ifc.cnr.it.

Abstract: The paper investigates the non-local property of quantum mechanics by using the quantum hydrodynamic analogy (QHA). The role of the quantum potential in generating the non-local dynamics of quantum mechanics is analyzed. The effectiveness of the Bohr-Sommerfeld quantization is investigated in the quantum hydrodynamic equations both in the deterministic as well as stochastic case. The work shows how in presence of noise the non-local properties as well as the quantization of the action are perturbed. The resulting quantum stochastic dynamics much depend by the strength of the interaction: Strongly bounded systems (such as linear ones) lead to a quantum entangled stochastic behavior, while weakly bounded ones may disentangle themselves and to be not able to maintain the quantum superposition of states on large distances acquiring the classical stochastic evolution .

The work shows that in the frame of the stochastic QHA approach it is possible to have *classical freedom* between quantum weakly bounded systems. The stochastic QHA model shows that the wave-function collapse to an eigenstates (deriving by interaction of a quantum microscopic system with the measuring apparatus in a classical environment) can be described by the model itself. Since, on the basis of the Copenhagen interpretation of quantum mechanics the time of the wave function decay to the eigenstate represents the minimum duration time of a measurement, the SQHA shows that the minimum uncertainty principle is compatible with the relativistic postulate about the light speed as the maximum velocity of transmission of interaction. The paper shows that the Lorenz invariance of the quantum potential enforces the hypothesis that superluminal transmission of information is not present in measurements on quantum entangled states.

## 1. Introduction

The quantum to classical transition is one of the most intriguing problems of the modern physics [1-3] The central point is the incompatibility between the non-local properties of quantum mechanics and the local ones of the classical dynamics [1-2]. The disconnection between the two theories leaves open the question about the hierarchy between them. The quantum mechanics, on the base of its statistical definition, needs the classical mechanics (i.e., the classical observer) to be defined, the quantum one seems to be the basic one from which the classical mechanics can stem out in the macroscopic limit where $\hbar$ tends to zero.

One current of thought is represented by the "deterministic" approach to quantum mechanics that analyzes how the quantum equations are a generalization of the classical one [4-13] where the nonlocality is introduced in various ways, the Madelung quantum potential [6-8], the Nelson's osmotic potential, the Bohm-Hylei quantum potential or the Paris and Wu fifth-time parameter.

As far as it concerns the QHA equations [6-8], the non-local restrictions come by applying the quantization of vortices [8] and by the elastic-like energy arising by the quantum pseudo-potential but not from boundary conditions. On the contrary, the Schrödinger equation is a differential equation where the non-local character of evolution is determined by the initial and boundary conditions that must be defined for describing a physical problem that are apart from the equation.

In the case of charged particles, the non-local properties of the Schrödinger equation come also from the presence of the electromagnetic (EM) potentials that depend by the intensities of EM fields in a non-local way (e.g., Aharonov –Bohm effect) [8].

In the corresponding hydrodynamic equations [4] the EM potentials do not appear but only in local way through the strength of the EM fields.

A more recent and sophisticated approach is given by t'Hooft [14-16]. It originates by outputs coming from the black-hole thermodynamics and by the so called holographic principle [17-18] that proposes the obtaining of the quantum mechanics through a process of loss of information .

The deterministic approach continuously gains interest in the physics community due to the fact that it helps in explaining quantum phenomena that cannot be easily described by the usual formalism. They are multiple tunneling [19], critical phenomena at zero temperature [20], mesoscopic physics [21-22], numerical solution of the time-dependent Schrödinger equation [23-25], quantum dispersive phenomena in semiconductors [26], quantum field theoretical regularization procedure [27] and the quantization of Gauge fields, without gauge fixing and without ensuing the Faddeev-Popov ghost [28].

On the theoretical point of view, one of the most promising aspect of these models is helping in investigating the quantum mechanical problems using efficient mathematical technique such as the stochastic calculus, the numerical approach and supersymmatry.

A parallel current of thought, investigates the possibility of obtaining the classical state through the loss of quantum coherence of classically chaotic systems due to the presence of stochastic fluctuations [30-33]. Most of the outputs of this field of investigation are based on numerical simulation and/or semi-empirical approach leaking of global theoretical view.

The present paper falls into the intersection of the two approaches: it investigates the non-local property of quantum mechanics and its decoherence as a consequence of fluctuations by using the QHA[6-8] implemented with the stochastic calculus. This strategy is supported by the advantage of QHA in managing the non-local quantum dynamics in system larger than a single atom when fluctuations becomes important [34-38].

The mathematically more clear statements of non-local restrictions of the QHA make it suitable for the achievement of the connection between quantum concepts (probabilities) and classical ones (e.g., trajectories) [39] helping in overcoming the contrast between the quantum behavior and the classical reality.

## 2. The quantum hydrodynamic analogy

The QHA-equations are based on the fact that the Schrödinger equation, applied to a wave function $\psi_{(q,t)} = |\psi|_{(q,t)} exp[\frac{i}{\hbar} S_{(q,t)}]$, is equivalent to the motion of a particle density $n_{(q,t)} = |\psi|^2_{(q,t)}$ with velocity $\dot{q} = \frac{\nabla S_{(q,t)}}{m}$, governed by the equations [8]

$$\partial_t n_{(q,t)} + \nabla \bullet (n_{(q,t)} \dot{q}) = 0, \tag{1}$$

$$\dot{q} = \nabla_p H = \frac{p}{m} = \frac{\nabla S_{(q,t)}}{m}, \tag{2}$$

$$\dot{p} = -\nabla(H + V_{qu}), \tag{3}$$

where

$$\nabla_p \equiv (\frac{\partial}{\partial p_1},....,\frac{\partial}{\partial p_{3n}}), \tag{4}$$

where

$$H = \frac{p \cdot p}{2m} + V_{(q)} \tag{5}$$

is the Hamiltonian of the system and where $V_{qu}$ is the quantum pseudo-potential that reads

$$V_{qu} = -(\frac{\hbar^2}{2m}) n^{-1/2} \nabla \cdot \nabla n^{1/2}. \tag{6}$$

Equation (1-3) with the identity

$$S = \int_{t_0}^{t} dt (\frac{p \cdot p}{2m} - V_{(q)} - V_{qu}) \tag{7}$$

can be derived [40-41] by the system of two coupled differential equations

$$\partial_t S_{(q,t)} = -V_{(q)} + \frac{\hbar^2}{2m} \frac{\nabla^2 |\psi|_{(q,t)}}{|\psi|_{(q,t)}} - \frac{1}{2m} (\nabla S_{(q,t)})^2 \tag{8}$$

$$\partial_t |\psi|_{(q,t)} = -\frac{1}{m} \nabla |\psi|_{(q,t)} \cdot \nabla S_{(q,t)} - \frac{1}{2m} |\psi| \nabla^2 S_{(q,t)} \tag{9}$$

by taking the gradient of (8) and multiplying equation (9) by $|\psi|$. It is straightforward to see that the system of equations (8-9) for the complex variable

$$\psi_{(q,t)} = |\psi|_{(q,t)} exp[\frac{i}{\hbar} S_{(q,t)}] \tag{11}$$

is equivalent to equate to zero the real and imaginary part of the Schrödinger equation

$$i\hbar \frac{\partial \psi}{\partial t} = -\frac{\hbar^2}{2m} \nabla^2 \psi + V_{(q)} \psi. \tag{12}$$

## 3. Stochastic generalization of the quantum hydrodynamic analogy

We consider here the presence of stochastic noise on the density $n_{(q,t)}$ as a function both of time and space. For the sufficiently general case, to be of practical interest, $\eta_{(q,t,T)}$ is assumed Gaussian with null correlation time, the space is assumed isotropic and the noises on different co-ordinates independent. Thence, the stochastic partial differential conservation equation for $n_{(q,t)}$ reads [38]

$$\partial_t n_{(q,t)} = -\nabla \cdot (n_{(q,t)} \dot{q}) + \eta_{(q,t,T)} \tag{13}$$

$$<\eta_{(q_\alpha,t)}, \eta_{(q_\beta+\lambda,t+\tau)}> = <\eta_{(q_\alpha)}, \eta_{(q_\beta)}> G(\lambda) \delta(\tau) \delta_{\alpha\beta} \tag{14}$$

$$\dot{p} = -\nabla (V_{(q)} + V_{qu(n)}), \tag{15}$$

$$\dot{q} = \frac{\nabla S}{m} = \frac{p}{m}, \tag{16}$$

$$S = \int_{t_0}^{t} dt \left( \frac{p \cdot p}{2m} - V_{(q)} - V_{qu(n)} \right) \tag{17}$$

where $T$ is the noise amplitude parameter (i.e., the temperature of an ideal gas thermostat in equilibrium with the system[38]) and $G(\lambda)$ is the dimensionless shape of the spatial correlation function of $\eta$.

The condition that the energy fluctuations due to the quantum potential $V_{qu(n)}$ do not diverge, as $T$ goes to zero (so that the deterministic limit (i.e., the quantum mechanics) can be warranted) leads to a $G(\lambda)$ owing the form [38]

$$\lim_{T \to 0} G(\lambda) = exp\left[ -\left( \frac{\lambda}{\lambda_c} \right)^2 \right]. \tag{18}$$

The noise spatial correlation function (18), is a direct consequence of the derivatives present into the quantum potential that give rise to an elastic-like contribution to the system energy that reads [41]

$$\overline{H}_{qu} = \int_{-\infty}^{\infty} n_{(q,t)} V_{qu(q,t)} dq = -\int_{-\infty}^{\infty} n_{(q,t)}^{1/2} \left( \frac{\hbar^2}{2m} \right) \nabla \cdot \nabla n_{(q,t)}^{1/2} dq, \tag{19}$$

where a large "curvature" of $n_{(q,t)}$ leads to high quantum potential energy (see figure 1).

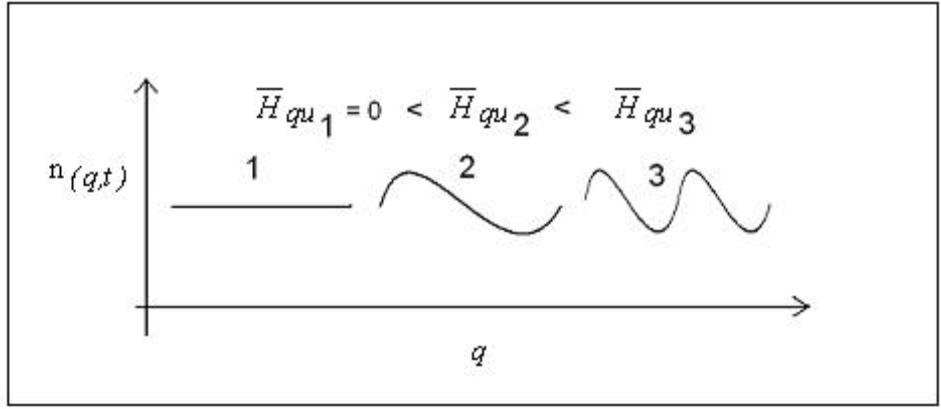

Figure 1  The quantum potential energy as a function of the "curvature" of the particle density n

This can be easily checked by calculating the quantum potential of the wave function $\psi = cos\frac{2\pi}{\lambda}q$ that reads

$$V_{qu} = -\left( \frac{\hbar^2}{2m} \right) \left( cos^2 \frac{2\pi}{\lambda} q \right)^{-1/2} \nabla \cdot \nabla \left( cos^2 \frac{2\pi}{\lambda} q \right)^{1/2} = \frac{\hbar^2}{2m} \left( \frac{2\pi}{\lambda} \right)^2 \tag{20}$$

showing that the energy increases as the inverse squared of the distance $\lambda$ between two adjacent peaks (i.e, the wave length). In the stochastic case, these peaks can be generated by two independent fluctuations of the wave function modulus where $\lambda$ represents (as a mean) the correlation distance of such fluctuations.

Therefore, particle density independent fluctuations very close each other (i.e., $\lambda \rightarrow 0$), generating very high curvature on the density $n_{(q,t)}$, can lead to a whatever large quantum potential energy even in the case of vanishing fluctuations amplitude (i.e., $T \rightarrow 0$).

In this case the convergence of equations (13-17) to the deterministic limit (1-3) (i.e., the standard quantum mechanics) would not happen. Therefore, in order to eliminating these unphysical solutions, the additional conditions (22) comes into the set of the quantum equations [38] in order to rule out unphysical solutions.

If we require that $\overline{H}_{qu} < \infty$ (following the criterion that higher is the energy lower is the probability to reach the corresponding (i.e., state with infinite energy have zero probability to realize itself) it follows that independent fluctuations of the density $n_{(q,t)}$ on shorter and shorter distance are progressively suppressed (i.e., have lower and lower probability of happening). This physical effect due to the quantum potential (that confers to the particle density function the elastic behavior like a membrane, very rigid against short range curvature) imposes a finite correlation length to the possible physical fluctuations.

In the small noise limit [38] the suppression of particle density fluctuations on very short distance, due to the finite energy requirement, brings to a restriction on the correlation length of the noise itself $\lambda_c$ [38] that reads

$$\lim_{T \rightarrow 0} \lambda_c = (\frac{\pi}{2})^{3/2} \frac{\hbar}{(2mkT)^{1/2}}, \qquad (21.a)$$

and to the expression for $G(\lambda)$ that reads

$$\lim_{T \rightarrow 0} G(\lambda) = exp[-(\frac{\lambda}{\lambda_c})^2] \qquad (21.b)$$

leading to explicit form of the variance (18)

$$\lim_{T \rightarrow 0} <\eta_{(q_\alpha,t)}, \eta_{(q_\beta+\lambda,t+\tau)}> = \mu \frac{kT}{2\lambda_c^2} exp[-(\frac{\lambda}{\lambda_c})^2]\delta(\tau)\delta_{\alpha\beta} \qquad (21.c)$$

where $\mu$ is a constant with the dimension of a migration coefficient.

Furthermore, the action (17), that can be re-cast in th form [38]

$$S = \int_{t_0}^{t} dt (\frac{p \cdot p}{2m} - V_{(q)} - V_{qu(n)})$$

$$= \int_{t_0}^{t} dt (\frac{p \cdot p}{2m} - V_{(q)} - V_{qu(n_0)} - \delta V_{qu}) \qquad , \qquad (22)$$

$$= S_0 + \delta S$$

in the case of very small noise amplitude (close to the deterministic quantum mechanical limit) due to the constraints (21.c), owns a $\delta S$ that is a small fluctuating quantity[38].

Finally, it is worth mentioning that for $T > 0$ the stochastic equations (13-17) can be obtained by the following system of differential equations

$$\partial_t S_{(q,t)} = -V_{(q)} + \frac{\hbar^2}{2m} \frac{\nabla_q^2 A_{(q,t)}}{A_{(q,t)}} - \frac{1}{2m}(\nabla_q S_{(q,t)})^2 \qquad (23)$$

$$\partial_t A_{(q,t)} = -\frac{1}{m}\nabla A_{(q,t)} \bullet \nabla S_{(q,t)} - \frac{1}{2m}A\nabla^2 S_{(q,t)} + A^{-1}\eta_{(q,t,T)} \qquad (24)$$

which for the complex wave function $\psi$ are equivalent to following the stochastic version of the Schrödinger equation [40]

$$i\hbar\frac{\partial \psi}{\partial t} = -\frac{\hbar^2}{2m}\nabla^2\psi + V_{(q)}\psi + i\frac{\psi}{|\psi|^2}\eta_{(q,t,T)}. \qquad (25)$$

## 4. Analysis of quantization condition and quantum coherence in presence of stochastic noise

If we look at the manageability of the quantum equations no one would solve the hydrodynamic ones. Nevertheless, the interest for the QHA remained unaltered along the time. The motivation for this does not only reside in the formal analogy with the classical mechanics, but also in the fact that the non-local properties of quantum mechanics (deriving by the quantization conditions) are more clearly mathematically recognizable in the QHA model.

In order to establish the hydrodynamic analogy, the gradient of equation (8,23) is considered. When we do that, we broaden the solutions of (12,25) so that not all solutions of the hydrodynamic equations can be solutions of the Schrödinger problem.

As well described in ref. [8], the state of a particle in the QHA is defined by the real functions $n_{(q,t)}$ and

$$p = \nabla S_{(q,t)}.$$

The restriction of the class of solutions of the QHA problem comes from additional conditions deriving from the quantization condition on the action. The integrability of the action gradient, in order to warrant the existence of the scalar action function, is warranted if the probability fluid is irrotational, that being

$$S_{(q,t)} = \int_{q_0}^{q} dl \bullet p \qquad (26)$$

it is to say that

$$\nabla \times p = 0 \qquad (27)$$

and hence that

$$\Gamma c = \oint dl \bullet m\dot{q} = 0 \qquad (28)$$

Moreover, since the action is contained in the argument of the exponential function of the wave function, all the multiples of $2\pi\hbar$, with $\Gamma c = 0 \pm 2n\pi\hbar$ on a closed contour, are accepted.

In the QHA, these non-local characteristics are transferred to the dynamics through the quantum potential (6) to which the quantized action is linked by formula (22). This is confirmed by the fact that if the contribution of the quantum potential is subtracted from the quantum equation, the classic non-linear Schrödinger one is obtained [40]. On the other hand, if the quantum potential is null, the hydrodynamic equations describe the motion of a classical dust of density $n_{(q,t)}$.

In the Schrödinger problem not all solutions are retained, but only those that fulfill precise boundary conditions (e.g., for bounded problems the eigenstates are those that goes to zero at infinity).

In the QHA the eigenstates are defined by the stationarity that happens when the force generated by the quantum potential exactly counterbalance that one due to the Hamiltonian potential (with the initial condition $\dot{q} = 0$).

Since the quantum potential changes with the state of the system, more than one stationary state (each one with its own $V_{qu}{}^n$) is possible and more than one quantized eigenvalues of the energy may exist with the corresponding action values

$$S_0{}^n = \int_{t_0}^{t} dt(\frac{p \cdot p}{2m} - V_{(q)} - V_{qu_0}{}^n) \qquad (29)$$

The above statements can be straightforwardly checked in the case of a linear system. For a harmonic oscillator described by the Hamiltonian $H = \frac{p^2}{2m} + \frac{m\omega^2}{2}q^2$, whose generic $n$-th eigenstate reads

$$\psi_{n(q)} = n^{\frac{1}{2}}{}_{(q,\,t)} \exp[\frac{i}{\hbar}S_{(q,t)}] = H_n(\frac{m\omega}{2\hbar}q)\exp\left(-\frac{m\omega}{2\hbar}q^2\right), \qquad (30)$$

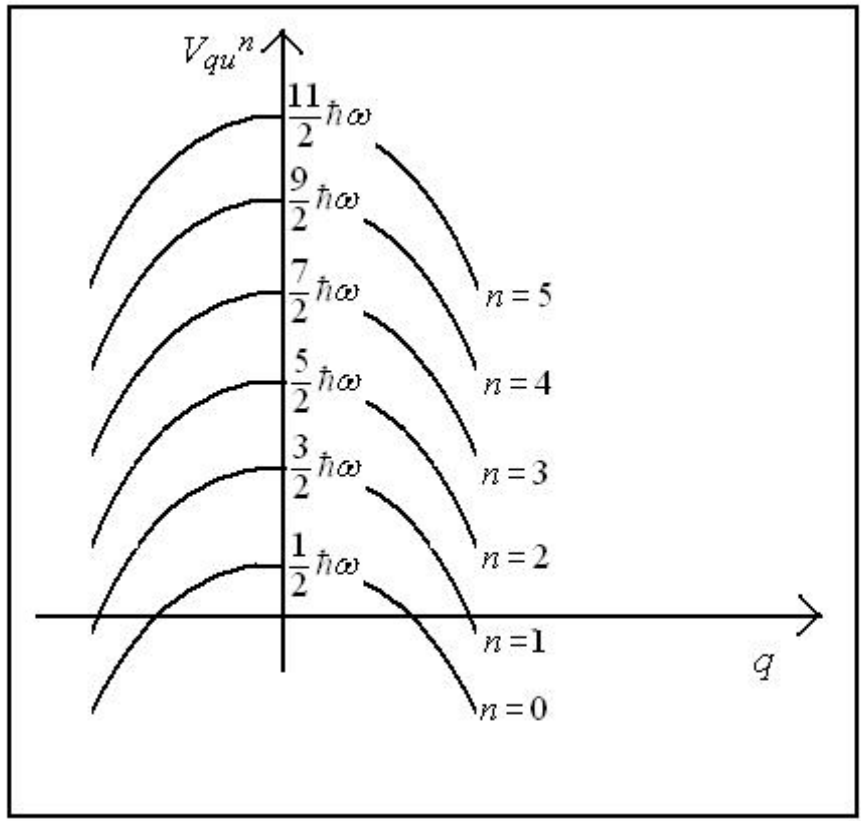

Figure 2. The repulsive quantum potential for the firsts five eigenstates of a harmonic oscillator

(where $H_{n(x)}$ represents the $n$-th Hermite polynomial) the density $n_{(q,\,t)}$ and the action $S_{(q,t)}$ respectively read

$$n^{\frac{1}{2}}{}_{(q,\,t)} = H_n(\frac{m\omega}{2\hbar}q)\exp\left(-\frac{m\omega}{2\hbar}q^2\right) \qquad (31)$$

$$S_{(q,t)} = S_{(t)}, \qquad (32)$$

leading to the quantum potential of the $n$-th eigenstate (see figure 2)

$$V_{qu}{}^n = -(\frac{\hbar^2}{2m})n^{\frac{1}{2}}{}_{(q)}\nabla_q \cdot \nabla_q n^{\frac{1}{2}}{}_{(q)}$$

$$= -\frac{m\omega^2}{2}q^2 + \left[n\left(\frac{\frac{m\omega}{\hbar}H_{n-1} - 2(n-1)H_{n-2}}{H_n}\right) + \frac{1}{2}\right]\hbar\omega \qquad (33)$$

$$= -\frac{m\omega^2}{2}q^2 + (n+\frac{1}{2})\hbar\omega$$

where it has been used the recurrence formula of the Hermite polynomials

$$H_{n+1} = \frac{m\omega}{\hbar}qH_n - 2nH_{n-1}, \qquad (34)$$

that gives the following energy eigenvalues

$$E_n = <\psi_n | H | \psi_n> = \int_{-\infty}^{\infty} n_{(q,t)}\left[\frac{m\omega^2}{2}(q-\underline{q})^2 + V_{qu}{}^n + \frac{m}{2}|\overset{*}{q}|^2\right]$$

$$= \int_{-\infty}^{\infty} n_{(q,t)}\left[\frac{m\omega^2}{2}(q-\underline{q})^2 + V_{qu}{}^n + \frac{1}{2m}|\nabla S_{(q)}|^2\right] \qquad (35)$$

$$= \int_{-\infty}^{\infty} n_{(q,t)}\left[\frac{m\omega^2}{2}(q-\underline{q})^2 - \frac{m\omega^2}{2}(q-\underline{q})^2 + (n+\frac{1}{2})\hbar\omega\right] = (n+\frac{1}{2})\hbar\omega$$

as well as

$$\overset{\bullet}{p} = -\nabla(H + V_{qu}) = -\nabla((n+\frac{1}{2})\hbar\omega) = 0, \qquad (36)$$

$$\overset{\bullet}{q} = \frac{\nabla S_{(q,t)}}{m} = 0, \qquad (37)$$

$$S_0{}^n = \int_{t_0}^{t} dt(\frac{p \cdot p}{2m} - V_{(q)} - V_{qu_0}{}^n) = E_n(t - t_0) \qquad (38)$$

In the QHA, the non-locality does not come from boundary conditions (that are apart from the equations) but from the quantum pseudo-potential (6) that depends by the state of the system and is a source of an elastic-like energy [8, 38, 41].
If we consider a bi-dimensional space, the quantum potential makes the vacuum acting like an elastic membrane that becomes quite rigid against curvature (i.e., fluctuations) on very small distances.
Given that the force of the quantum potential in a point depends by the state of the system around it, it introduces the non-local character into the motion equations.
Being so, the quantum non-local properties can be very well identified and studied by means of the analytical mathematical investigations of the property of the quantum potential (6).
This fact is even more important in presence of fluctuations since the quantum potential, containing the second partial derivatives of the wave function modulus, is critically dependent by the distance on which independent fluctuations happen.

The derivation of the correlation length of the noise $\lambda_c$ from the condition of non-diverging energy of the quantum potential short-distance fluctuations brings a quite heavy stochastic calculation and is out of the purpose of this paper [38].

Nevertheless, from the general point of view, we can observe that if $\lambda_c$ goes to infinity respect to the physical length of the system $L$ (i.e., microscopic mass or low temperature) the noise variance (14) becomes a pure function of time and reads

$$\lim_{T \to 0} <\eta(q_\alpha,t),\eta(q_\beta+\lambda,t+\tau)> \cong \mu \frac{kT}{2\lambda_c^2}\delta(\tau)\delta_{\alpha\beta} \quad . \tag{39}$$

Moreover, given the $\lambda_c^{-2}$ dependence of the amplitude of noise variance, it goes to zero (as well as $\delta S$ in (22)) and the deterministic standard quantum equations are recovered in the limit $\lambda_c \to \infty$ [38].

In this case, the ensemble of forbidden values for the action due to the quantization constraint (deterministic limit) reaches the well-known characteristic of the quantum mechanics form (figure 3).

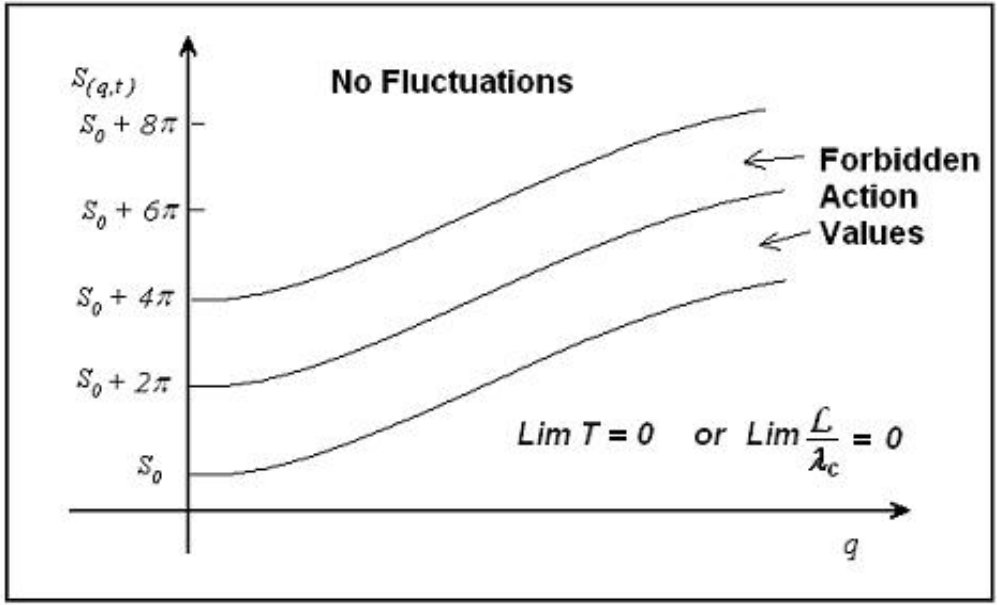

Figure 3. Quantized action in the case of a system without fluctuations

On the contrary, when fluctuations are present, the stochastic quantum hydrodynamic analogy leads to a scenario where the domain of forbidden states becomes smaller and smaller as the fluctuations amplitude increases (see figure 4))

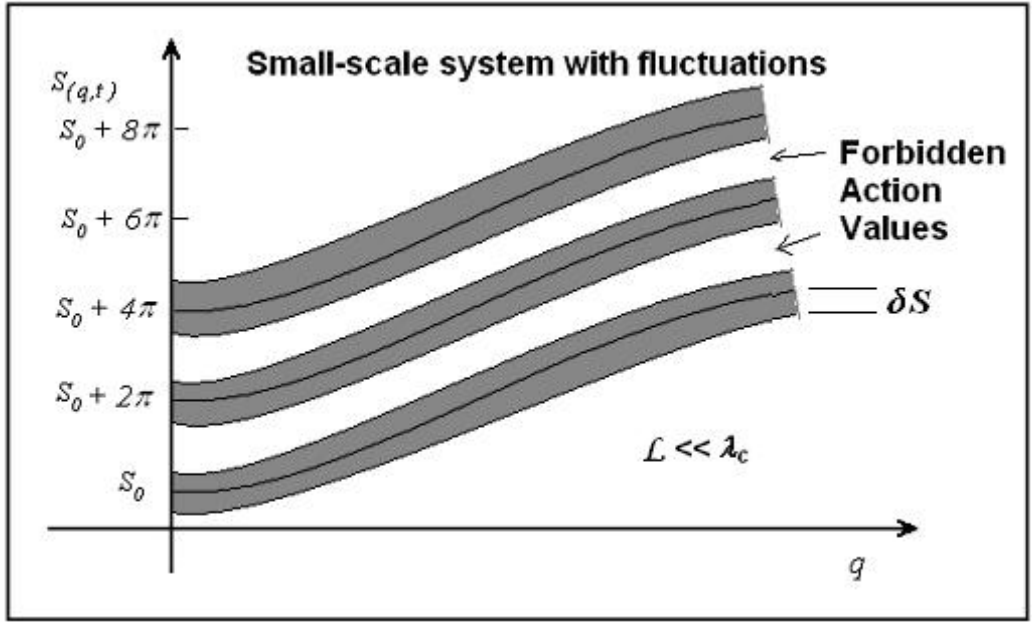

Figure 4. Quantized action in the case of a small-scale system submitted to small noise

In order to analytically detail what happens in the macroscopic case, mathematically speaking, we observe that the quantum force (equal to minus the gradient of the quantum potential) cannot be taken out by the deterministic PDE (1) (as intuitively proposed by many authors [8,41]) because this operation will wipe out the quantum stationary states (i.e., eigenstates) deeply changing the structure of such equation.

The presence of the QP is needed for the realization of the quantum stationary states (i.e., eigenstates) that happen when the force of the QP exactly balances the Hamiltonian one.

On the other hand, when we deal with large-scale systems with physical length $\mathcal{L} \gg \lambda_c$ and when fluctuations are present in weakly interacting systems, we can have a vanishing small quantum force at large distances (see appendix A) [22, 37, 38] that, becoming much smaller than fluctuations, can be correctly neglected in the motion equations.

It must be underlined that not all types of interactions lead to a vanishing small quantum force at large distance (a straightforward example is given by linear systems where the quantum potential owns a quadratic form (see appendix A) [22,38].

Nevertheless, it exists a large number of non-linear long-range weak potentials (e.g., Lennard Jones types) where the quantum potential tends to zero (see appendix A) at infinity and can be neglected [37]. In this case, a rarefied gas of such interacting particles behaves as a classical phase when the mean particle distance is much larger that the quantum potential range of interaction [22,37,38].

In the following we analyze the large scale form of the SPDE (13) both for finite and infinite quantum potential range of interaction.

In order to investigate this point, let's consider a system whose Hamiltonian reads

$$H = \frac{p^2}{2m} + V_{(q)}, \qquad (40)$$

in this case the equations (1-3) can be derived by the following phase-space equation

$$\partial_t \rho_{(q,p,t)} + \nabla \bullet (\rho_{(q,p,t)}(\dot{x}_H + \dot{x}_{qu})) = 0 \qquad (41)$$

where

$$n_{(q,t)} = \iiint \rho_{(q,p,t)} d^{3n}p . \qquad (42)$$

$$\overset{\bullet}{x}_H = (\partial_p H, -\nabla H) \qquad (43)$$

$$\overset{\bullet}{x}_{qu} = (0, -\nabla V_{qu}) \qquad (44)$$

by integrating equation (41) over the momentum $p$ with the conditions that $\lim_{|p|\to\infty} \rho_{(q,p,t)} = 0$ with the constraint on the quantum phase space density

$$\rho_{(q,p,t)} = n_{(q,t)} \delta(p - \nabla S) \qquad (45)$$

The factor $\delta(p - \nabla S)$ (namely the wave-particle equivalence) warrants the correspondence rule

$$p = m\overset{\bullet}{q} = \nabla S \qquad (46)$$

between the quantum hydrodynamic model and the Schrödinger equation [8,41].

When a spatially distributed random noise is present, the phase SPDE, whose zero noise limit is the deterministic PDE (41), reads

$$\partial_t \rho_{(q,p,t)} + \nabla \cdot (\rho_{(q,p,t)}(\overset{\bullet}{x}_H + \overset{\bullet}{x}_{qu})) = \eta_{(q,t,T)} \delta(p - \nabla S), \qquad (47)$$

Near the deterministic limit, in the case of Gaussian noise (8), it is possible to re-cast (47) as

$$\partial_t \rho_{(q,p,t)} + \nabla \cdot (\rho_{(q,p,t)}(\overset{\bullet}{x}_H + \overset{\bullet}{x}_{qu(\rho_0)})) = -\nabla \cdot (\rho_{(q,p,t)} \delta \overset{\bullet}{x}_{qu}) + \eta_{(q,t,T)} \delta(p - \nabla S), \qquad (48)$$

where $\rho_0$ in $\overset{\bullet}{x}_{qu(\rho_0)}$ is the solution of the PDE (1) and where $\delta \overset{\bullet}{x}_{qu} = (0, -\nabla \delta V_{qu})$, where

$$\delta V_{qu} = -(\frac{\hbar^2}{2m})\{n^{-1/2} \nabla \cdot \nabla n^{1/2} - n_0^{-1/2} \nabla \cdot \nabla n_0^{1/2}\}$$
$$= V_{qu(n)} - V_{qu(n_0)} \qquad (49)$$

where $n_{0(q,t)} = \iiint \rho_{0(q,p,t)} d^{3n}p$.

Thanks to conditions (21.a-21.b) [38], closer and closer we get to the deterministic limit (i.e., $\frac{\lambda_c}{\mathcal{L}} \to \infty$), smaller and smaller is the amplitude of the random term on the right side of (48)

$$-\nabla \cdot (\rho_{(q,p,t)} \delta \overset{\bullet}{x}_{qu}) + \eta_{(q,t,T)} \delta(p - \nabla S) = \xi_{(q,t)} \qquad (50)$$

When $\frac{\lambda_c}{\mathcal{L}} \to \infty$ the standard quantum mechanics is achieved and the quantum potential cannot be disregarded from the hydrodynamic quantum motion equations.

On the contrary, when $\lambda_c \ll \mathcal{L}$, in weakly bounded system when the force steaming from the quantum potential at large distance tends to zero (and becomes much smaller than its fluctuations) it is possible to coherently define [38] a measure of the quantum potential range of interaction $\lambda_q$ (see appendix A) that reads

$$\lambda_q = 2\frac{\int_0^\infty |q^{-1}\frac{\partial V_{qu}}{\partial q}|dq}{\lambda_c^{-1}|\frac{\partial V_{qu}}{\partial q}|_{(q=\lambda_c)}}, \qquad (51)$$

(For charged particles with spin (see Appendix B), $\lambda_q$ is determined by the Pauli's equation [8].

Thence, when $\frac{\lambda_q}{L} \to 0$ it follows that

$$\dot{x}_{qu(\rho_0)} \ll \delta \dot{x}_{qu} \qquad (52)$$

where formula (52) expresses the fact that the quantum potential force $\dot{x}_{qu(\rho_0)} = (0, -\nabla V_{qu(\rho_0)})$ is much smaller than its fluctuations $\delta \dot{x}_{qu(\rho_0)} = (0, -\nabla \delta V_{qu})$ and, hence, that

$$|\nabla V_{qu(n_0)}| \ll |\nabla \delta V_{qu(n)}|. \qquad (53)$$

For sake of completeness, we observe that close to the deterministic limit (i.e., to the quantum mechanics) when $L < \lambda_c$ the quantum potential cannot be disregarded even if it is vanishing small, therefore the quantum potential range of interaction $\lambda_q$ is physically meaningful if and only if $\lambda_q > \lambda_c$. For $\lambda_q < \lambda_c$ the quantum potential range of interaction must be retained equal to $\lambda_c$.

Introducing (53) into equation (48) it follows that

$$\partial_t \rho_{(q,p,t)} + \nabla \bullet (\rho_{(q,p,t)}(\dot{x}_H)) \cong -\nabla \bullet (\rho_{(q,p,t)} \delta \dot{x}_{qu}) + \eta_{(q,t,T)}\delta(p - \nabla S)$$

$$\lim_{\frac{\lambda_c}{L} \to 0} <\eta_{(q_\alpha,t)},\eta_{(q_\beta+\lambda,t+\tau)}> = \mu(\frac{kT}{2m})^{\frac{1}{2}}\delta(\lambda)\delta(\tau)\delta_{\alpha\beta} \qquad (L \gg \lambda_q) \quad (54)$$

Equation (54) for small but not null noise amplitude $T$ (i.e., $T \gg T_c = \frac{\hbar^2}{2mkL^2} \approx 3°K$ for $L = 3\times 10^{-8}$ m and $m$ equal to the proton mass, where $T_c$ is defined by setting $\lambda_c = L$) leads to the stochastic phase space PDE

$$\partial_t \rho_{(q,p,t)} + \nabla \bullet (\rho_{(q,p,t)}(\dot{x}_H)) = \xi'_{(q,t)} \qquad (55)$$

where $\xi'_{(q,t)}$ is a small random quantity, that shows dynamics that fluctuate around a deterministic "classical" core and that do not own eigenstates.

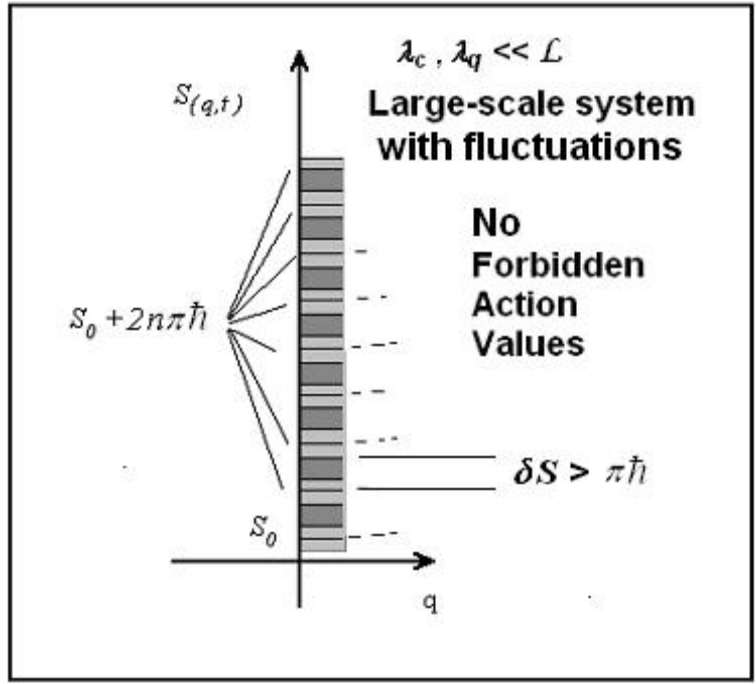

Figure 5. Quantized action in the case of a large-scale system (macroscopic mass or high temperature) submitted to small noise

Physically speaking, the central point in weakly quantum entangled systems, whose characteristic length is much bigger than the quantum potential range of interaction, is that the stochastic sequence of fluctuations of the quantum potential does not allow the coherent reconstruction of the superposition of state since they are much bigger than the quantum potential itself. In this case (especially in classically chaotic systems) the effect of the quantum potential with fluctuations (even with null time mean) on the dynamics of the system is not equal to the effect of its average.

If the quantum potential can be disregarded in a large scale description, the action (22) reads

$$S = \int_{t_0}^{t} dt(\frac{p \cdot p}{2m} - V_{(q)} - V_{qu(n_0)} - \delta V_{qu})$$

$$\cong \int_{t_0}^{t} dt(\frac{p \cdot p}{2m} - V_{(q)} - \delta V_{qu}) \qquad (56)$$

$$= S_{cl} + \delta S$$

and hence, the momentum of the solutions given by the $\delta$-function in (45) (i.e, $\delta(p - \nabla(S_{cl} + \delta S))$) approaches the classical value (plus a small fluctuation) and reads

$$p = \nabla(S_{cl} + \delta S) = p_{cl} + \delta p \qquad (57)$$

When we deal with a huge scale system (i.e., $\lambda_c << \mathcal{L}$) given the macroscopic scale, the quantized action values become very dense and the allowed action ones fill all the space (see figure 5). Moreover, in weakly interacting systems we may actually have that quantization is ineffective.

Observing that the quantum coherence length $\lambda_c$ results by the geometrical mean of the stochastic length $\frac{\hbar c}{kT}$ (of order of unity or less, about ($1,44\ cm$ at $1\ °K$)) and the Compton length $l_C = \frac{\hbar}{mc}$ (the reference length for the standard quantum mechanics) it follows that the description of a macroscopic system (with a resolution $\Delta q$ such as $\lambda_c, l_C, \lambda_q < \Delta q << \mathcal{L}$) is classically stochastic at laboratory scale, even at low temperature, since for $T_c$ as small as the temperature of the background radiation 2,725°K, it results

$$\lambda_c \propto (\frac{l_C \hbar c}{kT})^{1/2} = 2.8 \times 10^{-8}\ m$$ 

for a particle of proton mass (or $\lambda_c \approx 3 \times 10^{-9}\ m$ at a temperature of 300°K)). Even if the condition $\lambda_c < \lambda_q < \Delta q$ is usually satisfied for macroscopic objects constituted by Lennard-Jones interacting particles, there also exists (at laboratory condition) the possibility to have $\Delta q < \lambda_q$ and, hence, to detect quantum phenomena. The most direct and immediate way is to consider observables depending by molecular properties of solid crystals that, due to the linearity of the particles interaction, can own a very large quantum potential range of action $\lambda_q$ (that may result of order of ten times of the atomic distances [22]). Another possibility is to refrigerate a fluid below the critical density (if it does not undergo solidification) in order to obtain that the mean molecular distance becomes smaller than $\lambda_p$ or $\lambda_c$ [37].

## 5. Discussion

Even if the linear systems are the most studied and known ones, those characterized by non-linear weak interactions, to which equation (54) can apply, are more wide-spread in nature.

For instance, equation (54) can apply to the case of a rarefied gas phase of Lennard-Jones potential interacting particles where the mean inter-particle distance is much bigger than $\lambda_q$ and $\lambda_c$ (for instance for the helium at room temperature it results $\lambda_c \approx \lambda_q \approx 10^{-7}\ cm$ [37]). In this case, the quantum superposition of states of molecules (or group of them) do not exist so that the macroscopic gas system behaves classically.

A deeper analysis [22], shows that the classical behavior of molecules of a real gas is maintained down to the density of liquids. On the contrary, due to the linearity of intermolecular forces in solid crystals, $\lambda_q$ becomes bigger than the mean inter-particle distance [22] and the quantum behavior of groups of atoms is maintained. Nevertheless, since the linear interaction of solids ends over a certain distance, the quantum behavior survives just in phenomena depending by the molecular scale (e.g., Braggs diffraction).

From this output, the stochastic quantum hydrodynamic model gives a realistic answer to the Schrödinger's cat enigma: the quantum macroscopic cat (made of weakly interacting particles like ordinary molecules) does not have any real correspondence in a noisy environment.

Furthermore, it is worth mentioning that in the classical macroscopic reality when we try to detect microscopic variables, below a certain point the wave dual properties of particles emerge.

If in the classical macroscopic reality the particle concept owns the characteristic that position and velocity are perceived as independent, on microscopic scale the wave property of the matter (e.g., the impossibility to interact just with a part of the system without entirely perturbing it) leads to the coupling between conjugated variables such as position and velocity.

The consequence of the scale-dependence of the quantum potential interaction, leads the classical perception of the reality until the resolution size $\Delta q$ is at least larger than the quantum coherence length $\lambda_c$.

Moreover, we observe that higher is the amplitude of the noise $T$, smaller is the length $\lambda_c$ and, hence, higher is the attainable degree of spatial precision within the classical scale.

On the other hand, higher is the amplitude of noise, higher are the fluctuations of the observable such as the variance of velocity and/or energy measurements.

It is straightforward to show that this mutual opposite effect on conjugated variables is basically the same one leading to the Heisenberg's principle of uncertainty.

In fact, by using the quantum stochastic hydrodynamic model, it is possible to derive the uncertainty relation between the time interval $\Delta t$ of a measurement and the related variance of the energy on a particle of mass $m$.

If on distances smaller than $\lambda_c$ any system behave in quantum mode (as a wave) so that any its subparts cannot be perturbed without disturbing all the entire system, it follows that the independence between the measuring apparatus and the measured system (*classical freedom*) requires that they must be far apart, at least, more than $\lambda_c$ and hence for the finite speed of propagation of interactions and information (*local relativistic causality* (LRC)) the measure process must last longer than the time $\tau = \frac{\lambda_c}{c}$.

Moreover, given that the noise $\eta_{(q,t,T)}$ in (13) in the small noise limit (i.e, $T$ sufficiently small) leads to Gaussian energy fluctuations [38], it follows that the mean value of the energy fluctuation for the particle is $\Delta E_{(T)} = (\frac{3}{2})^{1/2} kT \propto kT$ [42] and thence, in the non-relativistic limit ($mc^2 >> kT$) for a particle of mass $m$, the energy variance $\Delta E$ reads

$$\Delta E \approx (<(mc^2 + \Delta E_{(T)})^2 - (mc^2)^2>)^{1/2} \cong (<(mc^2)^2 + 2\Delta E_{(T)} - (mc^2)^2>)^{1/2}$$
$$\cong (2mc^2 <\Delta E_{(T)}>)^{1/2} \cong (2mc^2 kT)^{1/2} \tag{58}$$

from which it follows that [38,42]

$$\Delta E \Delta t > \Delta E \Delta \tau = \frac{(2mc^2 kT)^{1/2} \lambda_c}{c} = \frac{\hbar}{2}. \tag{59}$$

It is worth noting that the product $\Delta E \Delta t$ is constant since the growing of the energy variance with the square root of the temperature $\Delta E \approx (2mc^2 kT)^{1/2}$ is exactly compensated by the decrease of the minimum time of measurement

$$\tau \propto \frac{\hbar}{(2mc^2 kT)^{1/2}} \qquad (60)$$

furnishing an elegant physical explanation why the Eisenberg relations exist in term of a physical constant.

The same result is achieved if we derive the uncertainty relation between the position and the momentum of a particle of mass $m$.

If we measure the spatial position of a particle with a precision of $\Delta L > \lambda_c$ so that we do not perturb its quantum wave function (that, due to environmental fluctuations, is spontaneously localized on a spatial domain of order of $\lambda_c$) the variance $\Delta p$ of the modulus of its relativistic momentum $(p^\mu p_\mu)^{1/2} = mc$ due to the vacuum fluctuations reads

$$\Delta p \approx (<(mc + \frac{\Delta E_{(T)}}{c})^2 - (mc)^2>)^{1/2} \cong (<(mc)^2 + 2m\Delta E_{(T)} - (mc)^2>)^{1/2}$$
$$\cong (2m<\Delta E_{(T)}>)^{1/2} \cong (2mkT)^{1/2} \qquad (61)$$

leading to the uncertainty relation

$$\Delta L \Delta p > \lambda_c \Delta p = \lambda_c (2mkT)^{1/2} = \frac{h}{2} \qquad (62)$$

If we impose of measuring the spatial position with a precision $\Delta L < \lambda_c$, we have to localize the quantum state of the particle more than what spontaneously is.

Due to the increase of spatial confinement of the wave function, an increase of the quantum potential energy (due to the overall increase of curvature of $n^{1/2}$) as well as the variance of its fluctuations are generated. As a consequence of this, the final particle momentum variance $\Delta p$ increases.

Since the correlation between the wave-function localization and momentum variance are submitted to the properties of the Fourier transform relations (holding for any wave system) the uncertainty relations remain satisfied anyway we try to localize the wave function (either by environmental fluctuations or by physical means (i.e., external potentials)).

On the other hand, as far as we remain in the classical scale so that we do not explore space domains whose length is smaller than $\lambda_c$ and leave the wave-function to its spontaneous localization (due to the external noise) the momentum variance is determined just by the environmental fluctuations.

In the frame of the stochastic QHA (SQHA) the achievement of the classical mechanics (having the property, described by Einstein, Podolsky and Rosen [2], of physical reality, freedom and local relativistic causality) is achieved as a scale mediated effect.

As far as the resolution limit of classical description is much larger than the length over which the wave (quantum) properties of the matter can be detected, those classical concepts are not contradicted. When we deal with observables of microscopic systems, the quantum properties arise since the quantum potential (i.e., the wave property of the matter) comes into effect.

The SQHA shows that the *classical freedom* principle (independence between systems) and the *local relativistic causality* can be achieved and made compatible with the quantum mechanics in the frame of a unique theory.

The possibility of *classical freedom* comes from the fact that weakly bounded systems can disentangle themselves beyond the quantum coherence lengths $\lambda_c$ and $\lambda_q$.

Furthermore, the needs of *classical freedom* in performing a measurement makes the uncertainty principle compatible with the *local relativistic causality* [42].

Moreover, it is noteworthy to note that the quantum mechanics recovered as the deterministic limit of a stochastic theory, fulfills the philosophical need of determinism [1-3]. In the SQHA model the quantum mechanics represent the deterministic limit of a stochastic theory. In this picture, the deterministic limiting quantum distribution functions are a sort of "mechanical" distributions (not statistical) whose evolution is deterministic and well defined once the initial distributions and boundary conditions are defined. Under this light the hydrodynamic model gives its answer to the God-enigma: God does not play dice.

Moreover, in the SQHA, the wave-function collapse to an eigenstate (due to an interaction (i.e., measurement) in a classical fluctuating environment) is not described in a statistical way (out of the theory) but can be described by the theory itself as a kinetic process to a stationary state. This fact leads to a quantum theory with the conceptual property of a complete theory.

If we want to support the Copenhagen interpretation of quantum mechanics (which without solution of continuity agrees with the experimental evidence) so that the wave-function exists at all points of space simultaneously, in the frame of the SQHA it is sufficient to assume that particles are correlated each other until they are separated by a distance smaller than $\lambda_c$ (and still may present quantum correlations (stochastic influenced) until they are separated by a distance smaller than $\lambda_q$).

Following this prescription, if two particles are quantum entangled, when the measurement on one of the two is performed and the global wave-function collapses to an eigenstate, we are not in front of an instantaneous process, but in front of a kinetic evolution toward a stationary state (in a finite region of space) with its characteristic time $\tau_c$.

Thence, since the "quantum relaxation" interval of time $\tau_c$ (the wave function decay time) represents the minimum time of measurement, the compatibility with the *local relativistic causality* implies that $\tau_c > \tau$.

From experimental point of view, in order to demonstrate that the *local relativistic causality* (LRC) breaks down in quantum processes, it needs to demonstrated that the decoherence time $\tau_c$ is so short that the wave function collapse to the eigenstate is faster than the light to travel the distance $\mathcal{L}$ over which the quantum entangled state is localized (with $\mathcal{L} > \lambda_c$ and $\mathcal{L} < \lambda_q$). Therefore, it is sufficient to demonstrate that

$$\tau_c < \frac{\lambda_c}{c}$$

but, since the environmental energy fluctuations for the particle are given by (21.a), it follows that, in the SQHA model, the demonstration of the LRC breaking is equivalent to prove the violation of the Heisenberg's uncertainty principle.

Moreover, since in the SQHA model the measure process is basically performed in presence of a macroscopic (classical) object (observer), and since the quantum de-localized states on a length bigger that $\lambda_q$ are subject to wave-function decoherence, it follows that quantum correlation can detected in a system with physical length $L < \lambda_q$.

Furthermore, since the LRC remains satisfied if $\tau_c > \frac{L}{c}$ and since the measurement ends not before the wave function decay is completely achieved so that the measuring time $\tau_m$ results $\tau_m > \tau_c$ it follows that

$$\tau_m > \frac{L}{c} \tag{63}$$

Therefore, in order to demonstrate that the establishing of quantum correlations is due to superluminal communications [46-47] it must result that

$$\tau_m < \frac{L}{c} \tag{64}$$

Let' make an example. Given two entangled photons traveling in opposite direction in the state

$$|\psi> = \frac{1}{\sqrt{2}}|H,H> + e^{i\phi}|V,V> \tag{65}$$

(where H and V stand for horizontal and vertical polarization) that cross polarizers oriented in the same direction.

If we are mentally oriented to think that the state of the photon is defined once it crosses the polarizer, the assumption that the state of the photon is defined only after the measurement has taken place leads to the conclusion that the photon superposition state interacts with polarizer but it is still not fully collapsed neither to $|H>$ nor to $|V>$ until it is adsorbed by the polarizer or by the photon counter and the measurement is completely done.

In this experiment the measurement time $\tau_m$ starts at the arrival of the first entangled photon at its polarizer-photon counter system and ends when the other entangled one is detected at the second polarizer-photon counter system. This point must be carefully evaluated since the contemporary detection of the photons at the two photon-counters systems, placed at the same distance from the source, does not mean that the duration of the measurement process is null.

Given $L$ the physical distance between the two polarizer-photon counter systems, thence the superluminal communications are ascertained if it results $\tau_m < \frac{L}{c}$.

From the theoretical point of view the satisfaction of the Lorentz invariance of the quantum potential can enforce the hypothesis of compatibility between the LRC and the quantum nonlocality. To this end we inspect the relativistic expression of the quantum potential given by the relativistic hydrodynamic formulations [43-44].

The relativistic quantum potential allows verifying if the non-local interactions involved in the quantum mechanics propagate themselves compatibly with the postulate of the relativity about the invariance of light speed as the fastest way to which signals and interactions are transmitted.

Given the invariance of light speed is the generating property of the Lorentz transformations, the co-variant form (i.e., invariant 4-scalar product) of quantum potential of the Dirac equation that reads [43]

$$V_{qu} = -\frac{\hbar}{2i}\left[\overset{\bullet}{q}^{\mu} \partial_{\mu} \, ln[\frac{R}{\overline{R}}]\right] \qquad (66)$$

where

$$\overset{\bullet}{q}^{\mu} = \frac{\overline{\Psi} c \gamma^{\mu} \Psi}{|\Psi|^2} = \frac{\Psi^* c \gamma^0 \gamma^{\mu} \Psi}{|\Psi|^2} \qquad (67)$$

$$ln[\frac{R}{\overline{R}}] = \left(ln[\frac{|\Psi_1|}{|\Psi_3|}], \; ln[\frac{|\Psi_2|}{|\Psi_4|}], \; ln[\frac{|\Psi_3|}{|\Psi_1|}], \; ln[\frac{|\Psi_4|}{|\Psi_2|}]\right) \qquad (68)$$

and where $\Psi_i$ are the components of the 4-dimensional wave function

$$\Psi = (\Psi_1, \Psi_2, \Psi_3, \Psi_4),$$

united to the property of the 4-dimensional wave function $\Psi$ that changes accordingly with the Lorentz transformation, allows affirming that the quantum non-local behavior (deriving by the quantum potential) is compatible with such a postulate of the relativity.

In fact, whatever inertial system we choose moving with velocity $v < c$, the quantum potential expression (66) describes the quantum dynamics as realize themselves in such new reference system (where the light speed is always $c$ and hence not attainable). This fact forbids that in any inertial system the time difference between the initial conditions (e.g., starting of measurement (i.e., cause)) and the final one (wave collapse (i.e., effect)) is null (or negative) so that the quantum-potential action on the whole wave function (sometime de-localized on very far away points) cannot realize itself in a null time.

This result enforces the hypothesis that any measurable quantum non-local process (even involving a large distance) is compatible with the postulate of invariance of light speed as the fastest way to which signals and interactions can be transmitted.

It is a matter of fact that the compatibility between the quantum mechanics and the postulate of light speed invariance of the relativity needs the definition of a theory able to describe the kinetic of the wave function collapse during the measurement process.

Actually, the formulation of the standard quantum theory based on the statistical postulate concerning the measurement process makes it a semi empirical theory. On the other hand, a closed (self-standing) quantum theory must be able to describe the measuring process itself.

To this end the SQHA that has shown itself to be more suitable in describing phenomena at the edge between the quantum and classical mechanics (i.e., dispersive effects in semiconductors, critical phenomena, space geometry and gravitational models [45-48]) shows to be an effective candidate [12, 27].

# 7. Conclusion

In the present paper, the effect of the spatially distributed stochastic noise on the quantization is analyzed by implementing the quantum hydrodynamic analogy with the outputs of the stochastic calculus.

The work shows the quantum potential role in generating the non-local quantum behavior (eigenstates and coherent superposition of states) and its relationship with the quantization of velocity vortices leading the multiple quantized action values.

The analysis shows that in the frame of the quantum stochastic hydrodynamic model it is possible to maintain the concept of freedom of the classical reality between far away systems beyond the range of interaction of quantum potential as well as to make compatible the local relativistic causality with the uncertainty principle, one of the most relevant manifestations of the non-local behavior of the quantum mechanics.

In the frame of the SQHA, the wave-function collapse due to the interaction with a classical object (quantum disentangled system) submitted to environmental fluctuations, can be described inside the model itself so that it can be assimilated to a relaxation process to a stationary state with a finite proper time.

Compatibly with the Copenhagen approach of quantum mechanics, so that the wave-function collapse remains defined once the measurement process ends, the SQHA allows deriving the conditions on the measurement duration time in order to have the quantum measurements compatible with the relativistic postulate of invariance of light speed as the maximum one for the transmission of information.

The paper shows that this hypothesis has the theoretical support of the Lorentz invariance of the relativistic quantum potential that generates the nonlocal behavior of the quantum mechanics.

# References

1 .J.S. Bell "On the Einstain-Podolsky-Rosen paradox" Physics1, 195-200 (1964).

2. A. Einstein, B. Podolsky and N. Rosen, Phys, Rev., 47, 777-80 (1935).

3. Greenberger D.M., Horne, M. A., Shimony A., ZeilingerA., Bell's theorem without inequalities, Am. J. Phys. 58 (12) 1990, 1131-43.

4. I. Bialyniki-Birula, M., Cieplak, J., Kaminski, "Theory of Quanta", Oxford University press, Ny, (1992) 278.

5. D. Bohm, Phys Rev. 85 (1952) 166; D. Bohm, Phys Rev. 85 (1952) 180.

6. Madelung, E.:. Z. Phys. 40, 322-6 (1926).

7. Jánossy, L.: Zum hydrodynamischen Modell der Quantenmechanik. Z. Phys. 169, 79 (1962).8.

8. I. Bialyniki-Birula, M., Cieplak, J., Kaminski, "Theory of Quanta", Oxford University press, Ny, (1992), p. 87-111.

9. Nelson, E.: Phys. Rev. 150, 1079 (1966);

10. Nelson, E., *Dynamical Theory of Brownian Motion* (Princeton University Press, London, (1967);

11. Nelson, E., *Quantum Fluctuations* (Princeton University Press, New York, 1985);

12. F. Guerra, P. Ruggero, Phys rev. Lett. 31, 1022 (1973).

13. G. Parisi and Y.S. Wu, Sci. Sin. 24 (1981).

14. G. t'Hoft, J. Statistical Phys 53 (1988) 323.

15. G. t'Hoft, Quant. grav. 13 (1996) 1023.

16. G. t'Hoft, Class. Quant. grav. 16 (1999) 3263.

17. L. Susskind, L. Thorlacius, and J. Uglum, Phys. Rev. D 48 (1993) 3743.

18. R. Bousso, Rev. Mod Phys. 74 (2002) 825.

19. G. Jona, F. Martinelli, and E. Scoppola, Comm. Math. Phys. 80 (1981) 233.

20. Ruggiero P. and Zannetti, M., Phys. Rev. Lett. 47, 1231 (1981); Phys. Rev. Lett. 48, 963 (1982); Phys. Rev. B 27, 3001 (1983).

21. Ruggiero P. and Zannetti, M., Phys. Rev. A 28, 987 (1983).

22. Chiarelli, P., Quantum to classical transition in the stochastic hydrodynamic analogy: The explanation of the Lindemann relation and the analogies between the maximum of density at lambda point and that at the water-ice phase transition, Phys. Rev. & Res. Int. 3(4): 348-366, 2013.

23. Weiner, J. H. and Askar, A.: Particle Method for the Numerical Solution of the Time-Dependent Schrödinger Equation. J. Chem.
Phys. 54, 3534 (1971).

24. Weiner, J. H. and Forman, R.: Rate theory for solids. V. Quantum Brownian-motion model. Phys. Rev. B 10, 325 (1974).

25. Terlecki, G., Grun, N., Scheid, W.: Solution of the time-dependent Schrödinger equation with a trajectory method and application
to H+-H scattering. Physics Letters 88A, 33 (1982).

26. Gardner, C.L.: The quantum hydrodynamic model for semiconductor devices. SIAM J. Appl. Math. 54, 409 (1994).

27. J.D. Bret, S. Gupta, and A. Zaks, Nucl. Phys. B 233 (1984) 61.

28. D. Zwanziger, , Nucl. Phys. B 192 (1984) 259.

29. A. Mariano, P. Facchi, and S. Pascazio, Decoherence and Fluctuations in Quantum Interference Experiments, Fortschr. Phys. 49 (2001) 10—11, 1033 — 1039

30. Cerruti, N.R., Lakshminarayan, A., Lefebvre, T.H., Tomsovic, S.: Exploring phase space localization of chaotic eigenstates via parametric variation. Phys. Rev. E 63, 016208 (2000).

31. E. Calzetta and B. L. Hu, Quantum Fluctuations, Decoherence of the Mean Field, and Structure Formation in the Early Universe, Phys.Rev. D, **52**, 6770-6788, (1995).

32. C., Wang, P., Bonifacio, R., Bingham, J., T., Mendonca, Detection of quantum decoherence due to spacetime fluctuations, 37$^{th}$ COSPAR Scientific Assembly. Held 13-20 July 2008, in Montréal, Canada., p.3390.

33. F., C., Lombardo , P. I. Villar, Decoherence induced by zero-point fluctuations in quantum Brownian motion, Physics Letters A 336 (2005) 16–24

34. Tamura, H., Ramon, J. G. S., Bittner, E., Bourghardt, I., Phys Rev. Lett. 100,107402 (2008).

35. Bousquet D, Hughes KH, Micha DA Burghardt I. Extended hydrodynamic approach to quantum-classical nonequilibrium evolution I. Theory, J. Chem. Phys. 2001;134.

36. Morato L. M., Ugolini S, Stochastic description of a Bose–Einstein Condensate 2011;12(8):1601-1612.

37. P. Chiarelli, The quantum potential: the missing interaction in the density maximum of He$^4$ at the lambda point?, Am. J. Phys. Chem.. **2**(6) (2014) 122-131.

38. Chiarelli, P., Can fluctuating quantum states acquire the classical behavior on large scale? J. Adv. Phys. 2013; **2**, 139-163 .

39. Wyatt RE. Quantum dynamics with trajectories: Introduction to quantum hydrodynamics, Springer, Heidelberg; 2005.
40. Chiarelli, P., The classical mechanics from the quantum equation, Phys. Rew. & Res. Int. 2013;3(1):1-9.
41. Weiner, J.H., *Statistical Mechanics of Elasticity* (John Wiley & Sons, New York, 1983), p. 315-7.
42. Chiarelli, P., The Uncertainty Principle derived by the finite transmission speed of light and information J. Adv. Phys. 2013; 3 , 257-266.
43. Chiarelli, P., The quantum hydrodynamic formulation of Dirac equation and its generalized stochastic and non-linear analogs, accepted for publication to Phys. Sci. Int. J (2014).
44. Chiarelli, P., The relativistic quantum hydrodynamic representation of Klein-Gordon equation, *Quantum Matter*, 4, 1-7 (2015)
45. E. R. A. Giannetto, *Quantum Matter* 3, 273 (2014).
46. D. Fiscaletti, *Quantum Matter* 2, 45 (2013).
47. E. Koorambas, *Quantum Matter* 3, 215 (2014).
48. D. Fiscaletti, *Rev. Theor. Sci.* 1, 103 (2013).

# APPENDIX A

## Large-distance quantum force

To obtain the macro-scale form of equations (42-47) we need to evaluate the large-scale limit of the quantum force $\dot{p}_{qu} = -\nabla_q V_{qu}$ in (47). The behavior of $n^{1/2}$ determines the quantum potential (QP) in (6). For sake of simplicity, we discuss the one-dimensional case of localized state with $n^{1/2}$ that at large distance goes like

$$\lim_{|q|\to\infty} n^{1/2} \propto exp[-P^k(q)] \qquad (A.1)$$

where $P^k(q)$ is a polynomial of degree equal to $k$, $z_q = \gamma^{-1}q$ is the macroscopic variable (where $\gamma = \frac{\Delta q}{\lambda_q}$, where $\Delta q$ is the macro-scale resolution) and $\lambda_q$ is the range of the QP interaction. By using (A.1), the QP (6) at large scale reads

$$\lim_{\gamma\to\infty} V_{qu} = \lim_{\gamma\to\infty} -k^2 \gamma^{-\phi} z_q^{1-\phi} + k(k-1)\gamma^{-(1.5+\phi)} z_q^{-(3+\phi)/2} \qquad (A.2)$$

where $\phi = 3 - 2k$.

Thence, for $k < \frac{3}{2}$ (i.e., $\phi > 0$) $\forall\, z_q \neq 0$ finite, the quantum force $-\nabla_q V_{qu}$ at large scale (i.e., $\gamma \to \infty$, $q = \gamma z_q \to \infty$) reads

$$\lim_{\gamma \to \infty (z_q\, finite)} -\nabla V_{qu} = \lim_{q \to \infty} 2k^2(k-1)(\gamma z_q)^{-\phi} + k(k-1)(k-2)(\gamma z_q)^{-\frac{1}{2}(3+2\phi)} z_q^{\frac{1}{2}\phi}$$

$$\approx 2k^2(k-1)(\gamma z_q)^{-\phi} = 0 \qquad (A.3)$$

Moreover, since the integral

$$\int_0^\infty |q^{-1} \nabla V_{qu}|\, dq \cong Const + \alpha \int_{q_0}^\infty |\frac{1}{q^{1+\phi}}|\, dq < \infty \qquad (A.4)$$

converges for $\phi > 0$, (A.4) tells us if the QP force is negligible on large scale as given by (A.3). Therefore, finite values of the mean weighted distance

$$\lambda_q = 2 \frac{\int_0^\infty |q^{-1}\frac{\partial V_{qu}}{\partial q}|\, dq}{\lambda_c^{-1} |\frac{\partial V_{qu}}{\partial q}|_{(q=\lambda_c)}}, \qquad (A.5)$$

warrants the vanishing of QP at large distance and, hence, it can be assumed as an evaluation of the quantum potential range of interaction.

It is worth mentioning that condition (A.4) is not satisfied by linear systems whose eigenstates have $\phi = -1$ [29], so that $\lambda_q = \infty$ and they cannot admit the classical limit.

It is also worth noting that condition (A.4), obtained for $n^{1/2}$ (WFM) owing the form (A.1), also holds in the case of oscillating wave functions whose modulus is of type

$$\lim_{|q| \to \infty} n^{1/2} = |q^m \sum_n a_n \exp[iA_n^p(q)]\} \exp[-P^k(q)]| \qquad (A.6)$$

where $A_n^p(q)$ are polynomials of degree equal to $p$. In this case, in addition to the requisite $0 \leq k < \frac{3}{2}$, the conditions $m \in \Re$ and $p \leq 1$ are required to warrant (A.4) [38].

For instance, the Lennard-Jones-type potentials holds $\lim_{|q|\to\infty} A_n^p(q)_{(q)} \propto q$ and, hence, they own $\lambda_q$ finite.

In the multidimensional case, $\lambda_q$ depends by the path of integration $\Sigma$ and (A.5) reads

$$\lambda_q = 2 \frac{\int_{\Sigma_{r=0}}^{\Sigma_{r=\infty}} r_{(\Sigma)}^{-1} \left| \frac{\partial V_{qu}}{\partial q_i} \right| \cdot d\Sigma_i}{\lambda_c^{-1} \left| \frac{\partial V_{qu}}{\partial r} \right|_{(r_{(\Sigma)}=\lambda_c)}} \qquad (A.7)$$

where $r = |q|$ and $d\Sigma_i$ is the incremental vector tangent to $\Sigma$.

Since, the physical meaning of $\lambda_q$ must be independent by the path of integration (we know that $\frac{\partial V_{qu}}{\partial q_i}$ is integrable but do we do not know nothing about the integrability of $r_{(\Sigma)}^{-1} \left| \frac{\partial V_{qu}}{\partial q_i} \right|$) in order to well define $\lambda_q$ the fixation of the integral path is needed. If we choose the integration path $\Sigma = r m_i$ where $m_i$ is a generic versor, $\lambda_q$ reads

$$\lambda_{q(m_i)} = 2 \frac{\int_{r=0}^{\infty} r^{-1} \left| \frac{\partial V_{qu}}{\partial r} \right|_{(q=rm_i)} dr}{\lambda_c^{-1} \left| \frac{\partial V_{qu}}{\partial r} \right|_{(q=\lambda_c m_i)}} \qquad (A.8)$$

Moreover, since in order to evaluate at what distance the quantum force becomes negligible whatever is the direction of the versor $m_i$, among the values of (A.8) we must consider the maximum one so, finally, $\lambda_q$ reads

$$\lambda_q = Max \left\{ 2 \frac{\int_{r=0}^{\infty} r^{-1} \left| \frac{\partial V_{qu}}{\partial r} \right|_{(q=rm_i)} dr}{\lambda_c^{-1} \left| \frac{\partial V_{qu}}{\partial r} \right|_{(q=\lambda_c m_i)}} \right\} \qquad (A.9)$$

## Quantum potential characteristics

In order to elucidate the interplay between the Hamiltonian potential and the quantum potential, that together define the quantum evolution of the particle, we observe that the quantum potential is primarily defined by the WFM.

Fixed the WFM at the initial time, then the Hamiltonian potential and the quantum one determine the evolution of the WFM in the following instants that on its turn modifies the quantum potential.

A Gaussian WFM has a parabolic repulsive quantum potential, if the Hamiltonian potential is parabolic too (the free case is included), when the WFM wideness adjusts itself to produce a quantum potential that exactly compensates the force of the Hamiltonian one, the Gaussian states becomes stationary (eigenstates). In the free case, the stationary state is the flat Gaussian (with an infinite variance) so that any free Gaussian WFM expands itself following the ballistic dynamics of quantum mechanics since the Hamiltonian potential is null and the quantum one is a quadratic repulsive one.

From the general point of view, we can say that if the Hamiltonian potential grows faster than a harmonic one, the wave equation of a self-state is more localized than a Gaussian one and this leads to a stronger-than a quadratic quantum potential.

On the contrary, a Hamiltonian potential that grows slower than a harmonic one will produce a less localized WFM that decreases slower than the Gaussian one, so that the quantum potential is weaker than the quadratic one and it may lead to a finite quantum non-locality length (A.5).

More precisely, as shown above, the large distances exponential-decay of the WFM given by (A.1) with $k<3/2$ is a sufficient condition to have a finite quantum non-locality length [20].

In absence of noise, we can enucleate three typologies of quantum potential interactions (in the unidimensional case):

(1) $k > 2$ strong quantum potential that leads to quantum force that grows faster than linearly and $\lambda_q$ is infinite (*super-ballistic* expansion for the free particle WFM) and reads

$$\lim_{|q|\to\infty} \frac{\partial V_{qu}}{\partial q} \propto q^{1+\varepsilon}. \qquad (\varepsilon>0) \qquad (A.10)$$

(2) $k = 2$ that leads to quantum force that grows linearly

$$\lim_{|q|\to\infty} \frac{\partial V_{qu}}{\partial q} \propto q \qquad (A.11)$$

and $\lambda_q$ is infinite (*ballistic* expansion for the free particle WFM)

(3) $2 > k \geq 3/2$ "middle quantum potential";
the integrand of (A.4) will result

$$Cons\tan t > \lim_{|q|\to\infty} |q^{-1}\frac{\partial V_{qu}}{\partial q}| > q^{-1}. \qquad (A.12)$$

The quantum force remains finite or even becomes vanishing at large distance but $\lambda_q$ may be still infinite (*under-ballistic* expansion for the free particle WFM).

(4) $k < 3/2$ "week quantum potential" interaction leading to quantum force that becomes vanishing at large distance following the asymptotic behavior

$$\lim_{|q|\to\infty} |q^{-1}\frac{\partial V_{qu}}{\partial q}| > q^{-(1+\varepsilon)}, \varepsilon > 0 \qquad (A.13)$$

with a finite $\lambda_q$ for T ≠ 0 (*asymptotically vanishing* expansion for the free particle WFM).

## Pseudo-Gaussian particle

Gaussian particles generate a quadratic quantum potential that is not vanishing at large distance and hence cannot lead to macroscopic local dynamics. Nevertheless, imperceptible deviation by the perfect Gaussian WFM may possibly lead to finite quantum non-locality length. Particles that are inappreciably less localized than the Gaussian ones (let's name them as pseudo-Gaussian) own $\frac{\partial V_{qu}}{\partial q}$ that can sensibly deviate by the linearity so that the quantum non-locality length may be finite.

We have seen above that for $k < 3/2$ (when the WFM decreases slower than a Gaussian) a finite range of interaction of the quantum potential $\lambda_q$ is possible.

The Gaussian shape is a physically good description of particle localization, but irrelevant deviations from it, at large distance, are decisive to determine the quantum non-locality length.

For instance, let's consider the pseudo-Gaussian wave-function type

$$n = n_0 \, exp[-\frac{(q-\underline{q})^2}{\underline{\Delta q}^2[1+[\frac{(q-\underline{q})^2}{\Lambda^2 f(q-\underline{q})}]]}]\qquad(A.14)$$

where $f(q-\underline{q})$ is an opportune regular function obeying to the condition

$$\Lambda^2 f(0) \gg \underline{\Delta q}^2 \quad \text{and} \quad lim_{|q-\underline{q}|\to\infty} f(q-\underline{q}) \ll \frac{(q-\underline{q})^2}{\Lambda^2}.\qquad(A.15)$$

For small distance it holds

$$(q-\underline{q})^2 \ll \Lambda^2 f(q-\underline{q})\qquad(A.16)$$

and the localization given by the WFM is physically indistinguishable from a Gaussian one, while for large distance we obtain the behavior

$$lim_{|q-\underline{q}|\to\infty} n = n_0 \, exp[-\frac{\Lambda^2 f(q-\underline{q})}{\underline{\Delta q}^2}].\qquad(A.17)$$

For instance, we may consider the following examples

a)
$$f(q-\underline{q}) = 1\qquad(A.18)$$

$$lim_{|q-\underline{q}|\to\infty} n = n_0\, exp[-\frac{\Lambda^2}{\underline{\Delta q}^2}] ; \qquad (A.19)$$

b)
$$f(q-\underline{q}) = 1 + |q - \underline{q}| \qquad (A.20)$$

$$lim_{|q-\underline{q}|\to\infty} n = n_0\, exp[-\frac{\Lambda^2 |q-\underline{q}|}{\underline{\Delta q}^2}] ; \qquad (A.21)$$

c)
$$f(q-\underline{q}) = 1 + ln[1+|q-\underline{q}|^g] \approx ln[|q-\underline{q}|^g] \qquad (0<g<2) \qquad (A.22)$$

$$lim_{|q-\underline{q}|\to\infty} n \approx n_0\, |q-\underline{q}|^{-g\frac{\Lambda^2}{\underline{\Delta q}^2}} ; \qquad (A.23)$$

d)
$$f(q-\underline{q}) = 1 + |q-\underline{q}|^g \qquad (0<g<2) \qquad (A.24)$$

$$lim_{|q-\underline{q}|\to\infty} n = n_0\, exp[-\frac{\Lambda^2 |q-\underline{q}|^g}{\underline{\Delta q}^2}] \qquad (A.25)$$

All cases (a-d) lead to a finite quantum non-locality length $\lambda_q$.

In the case (d) the quantum potential for $|q-\underline{q}|\to\infty$ reads

$$lim_{|q-\underline{q}|\to\infty} V_{qu} = lim_{|q-\underline{q}|\to\infty} -(\frac{\hbar^2}{2m})|\psi|^{-1}\nabla_q\cdot\nabla_q|\psi|$$
$$= -(\frac{\hbar^2}{2m})[\frac{\Lambda^4 g^2 (q-\underline{q})^{2(h-1)}}{(2\underline{\Delta q}^2)^2} - \frac{\Lambda^2 g(g-1)(q-\underline{q})^{g-2}}{2\underline{\Delta q}^2}] \qquad (0<g<2) \qquad (A.26)$$

leading, for $0<g<2$, to the quantum force

$$lim_{|q-\underline{q}|\to\infty} -\nabla_q V_{qu} = (\frac{\hbar^2}{2m})[\frac{\Lambda^4 g^2(2g-1)(q-\underline{q})^{2g-3}}{(2\underline{\Delta q}^2)^2} - \frac{\Lambda^2 g(g-1)(g-2)(q-\underline{q})^{g-3}}{2\underline{\Delta q}^2}]$$
$$(A.27)$$

that for $g<3/2$ gives $lim_{|q-\underline{q}|\to\infty} -\nabla_q V_{qu} = 0$,

It is interesting to note that for $g=2$ (linear case)

$$|\psi| = n^{1/2} = n_0^{1/2} \exp[-\frac{(q-\underline{q})^2}{2\Delta q^2}] \tag{A.28}$$

the quantum potential is quadratic

$$\lim_{|q-\underline{q}|\to\infty} V_{qu} = -(\frac{\hbar^2}{2m})[\frac{(q-\underline{q})^2}{(\Delta q^2)^2} - \frac{1}{\Delta q^2}], \tag{A.29}$$

and the quantum force is linear (repulsive) and reads

$$\lim_{|q-\underline{q}|\to\infty} -\nabla_q V_{qu} = (\frac{\hbar^2}{2m})[\frac{2(q-\underline{q})}{(\Delta q^2)^2}] \tag{A.30}$$

The linear form of the force exerted by the quantum potential leads to the ballistic expansion (variance that grows linearly with time) of the free Gaussian quantum states.

# APPENDIX B

## Quantum correlation in particles with spin

When we are dealing with particles with spin (e.g., GHZ experiments [3]) instead of equations (1-3) we must consider the system of three equations describing the dynamics of spin waves and the interaction with the wave function density and the quantum moment (action gradient) that read

$$\partial_t n_{(q,t)} + \nabla \cdot (n_{(q,t)} \dot{q}) = 0, \tag{B.1}$$

$$\dot{q} = \partial_p H, \tag{B.2}$$

$$\left(\dot{p} - e\dot{A}\right)_i = \frac{d\dot{q}}{dt}_i = \frac{e}{m}\left(E + \dot{q}\times B\right)_i + \frac{\mu}{m} s_k \partial_i B_k + \frac{1}{n}\partial_j T_{ij}, \tag{B.3}$$

$$\dot{s}_i = \frac{2\mu}{\hbar}(\mathbf{s} \times \mathbf{B}) + \frac{1}{n}\partial_j N_{ij}, \tag{B.4}$$

where

$$T_{ij} = \frac{\hbar^2}{4m^2} n\left[\partial_i \partial_j \ln n + (\partial_i s_k)(\partial_j s_k)\right] \tag{B.5}$$

$$N_{ij} = \frac{\hbar}{2m} n \varepsilon_{ilm} s_l \partial_j s_m \tag{B.6}$$

$$s_i = \frac{\psi^\dagger \sigma_i \psi}{\psi^\dagger \psi} = \frac{\psi^\dagger \sigma_i \psi}{n} = \chi^\dagger \sigma_i \chi \tag{B.7}$$

where $\sigma_i$ are the Pauli matrices and $\psi = R_{(q,t)} \exp[i\frac{S_{(q,t)}}{\hbar}](\chi_{(q,t)})$ [40]. When we introduce the noise, equation (B.1) becomes

$$\partial_t n_{(q,t)} = -\nabla \cdot (n_{(q,t)} \dot{q}) + \eta_{(q,t,T)} \tag{B.8}$$

$$<\eta_{(q_\alpha,t)}, \eta_{(q_\beta+\lambda,t+\tau)}> = <\eta_{(q_\alpha)}, \eta_{(q_\beta)}> G(\frac{\lambda}{\lambda_c})\delta(\tau)\delta_{\alpha\beta} \tag{B.9}$$

where the form of correlation function $G(\frac{\lambda}{\lambda_c})$ is determined by imposing the condition that the mean squared root of the energy fluctuations does not diverge to infinity.

For a preliminary evaluation, we can take the hydrodynamic formula of the mean energy of the system [8]

$$<E> = \int d^3q \left[ n_{(q,t)} \frac{m\dot{q}^2}{2} + \frac{\hbar^2}{8m}\left(\frac{1}{n}(\nabla n)^2 + n\sum_i (\nabla s_i)^2\right)\right] \tag{B.10}$$

from which we can see that the dependence of the mean energy from the spin variables $\sum_i (\nabla s_i)^2$ is of the same type of that one $(\nabla n)^2$ of the particle density n.

Since the correlation function $G(\frac{\lambda}{\lambda_c})$ is determined by the order of the derivative of those terms [38], we expect that the correlation function $G(\frac{\lambda}{\lambda_c})$ as well as the correlation length $\lambda_c$ (of density and spin fluctuations) will be of the same type of (21.a-21.b).

As far as the quantum potential range of action $\lambda_q$, in the case of particle with spin, the additional terms $\frac{\hbar^2}{4m^2} n[(\partial_i s_k)(\partial_j s_k)]$ and $N_{ij} = \frac{\hbar}{2m} n \varepsilon_{ilm} s_l \partial_j s_m$ must be analyzed at large distance.

In order to evaluate when the spin quantum interactions can be disregarded (so that the classical scale sets up) we have to compare the gradient of the quantum potential of the standard quantum mechanics (i.e., deterministic limit) with its fluctuations when noise is present.

In order to evaluate if the quantum force (generating the quantum entanglement) has a finite range of action we preliminarily analyze the condition under which it results that

$$lim_{|q| \to \infty} \frac{\hbar^2}{4m^2} \frac{1}{n} \partial_i n[(\partial_i s_k)(\partial_j s_k)] = 0 \tag{B.11}$$

and that

$$lim_{|q| \to \infty} \frac{\hbar}{2m} \frac{1}{n} \partial_i n \varepsilon_{ilm} s_l \partial_j s_m = 0 \ . \tag{B.12}$$

By solving by parts the partial derivatives we have that

$$lim_{|q| \to \infty} \left( \partial_i [(\partial_i s_k)(\partial_j s_k)] + [(\partial_i s_k)(\partial_j s_k)] \partial_i ln[n] \right) = 0 \tag{B.13}$$

and that

$$lim_{|q| \to \infty} \left( \partial_i \varepsilon_{ilm} s_l \partial_j s_m + \varepsilon_{ilm} s_l \partial_j s_m \partial_i ln[n] \right) = 0 \tag{B.14}$$

where the first terms in both expressions do not depend by the particle density $n$.

The terms depending by $\partial_i ln[n]$ become vanishing at infinity for pseudo-Gaussian localization such as

$$n \approx exp[-q^{1-\varepsilon}] \tag{B.15}$$

with $\varepsilon > 0$.

As far as it concerns the terms

$$lim_{|q| \to \infty} \partial_i [(\partial_i s_k)(\partial_j s_k)] \tag{B.16}$$

and

$$lim_{|q|\to\infty} \partial_i \varepsilon_{ilm} s_l \partial_j s_m , \tag{B.17}$$

given the sine and cosine dependence upon the orientation of the spin vector $s_i$, in principle we may relevant quantum potential spin interaction at large distance. This result shows that the spin entanglement is the more suitable effect for the experimental study of quantum correlations on macroscopic distances.

**CAPTIONS**

Figure 1  The quantum potential energy as a function of the "curvature" of the particle density n

Figure 2. The repulsive quantum potential for the firsts five eigenstates of a harmonic oscillator

Figure 3. Quantized action in the case of a system without fluctuations

Figure 4. Quantized action in the case of a small-scale system submitted to small noise

Figure 5. Quantized action in the case of a large-scale system submitted to small noise